\documentclass[aps,showpacs,twocolumn]{revtex4-1}
\usepackage[english]{babel}
\usepackage{graphicx}
\usepackage{color}
\usepackage{times}
\usepackage{amssymb}

\begin{document}
\bibliographystyle{prsty}
\title{Taming L\'evy flights in confined crowded geometries}
\author{Micha\l{} Cie\'sla$^1$}
 \email{michal.ciesla@uj.edu.pl}
\author{Bart\l{}omiej Dybiec$^1$}
 \email{bartek@th.if.uj.edu.pl}
\author{Ewa Gudowska-Nowak$^1$}
 \email{gudowska@th.if.uj.edu.pl}
\author{Igor Sokolov$^2$}
 \email{igor.sokolov@physik.hu-berlin.de}
\affiliation{$^1$M. Smoluchowski Institute of Physics, Jagiellonian University, 30-348 Krak\'ow, \L{}ojasiewicza 11, Poland}%
\affiliation{$^2$Institut f\"ur Physik, Humboldt Universit\"at zu Berlin, Newtonstr. 15, 12489 Berlin, Germany}
\date{\today}
\begin{abstract}
We study a two-dimensional diffusive motion of a tracer particle in restricted, crowded anisotropic geometries. The underlying medium is the same as in our previous work [{\em J. Chem. Phys.} {\bf 140},  044706  (2014)]
in which standard, gaussian diffusion was studied. Here, a tracer is allowed to perform Cauchy random walk with uncorrelated steps. Our analysis shows that 
presence of obstacles significantly influences motion, which in an obstacle-free space would be of a superdiffusive type. At the same time, the selfdiffusive process reveals different anomalous properties, both at the level of a single trajectory realization and after the ensemble averaging. In particular, due to obstacles, the sample mean squared displacement asymptotically grows sublinearly in time, suggesting non-Markov character of motion. Closer inspection of survival probabilities indicates however that underlying diffusion is memoryless over long time scales despite strong inhomogeneity of motion induced by orientational ordering.
\end{abstract}
\maketitle
%
%
\section{Introduction\label{sec:introduction}}

Diffusive motion is a fundamental type of transport in biology - from the motion of animals to subcellular transport of organelles and chemicals in the crowded environment of the cell. Intense research on diffusive transport phenomena has been vitalized over past decades by progress in manufacturing nanoscale  molecular sieves and probing microrheological properties of porous materials. A standard mathematical tool to describe the transport of small particles suspended in simple solvents has been put forward by works of Einstein and Smoluchowski, both pointing to the probabilistic nature of performed motion in which displacements become random variables.   Pertinent to this approach is the concept of the continuous time random walks \cite{montroll1965,montroll1984,metzler2000} which describes diffusive motion  as a ``jump and wait'' process. In the simplest situations the waiting time and jump length of the walker can be fixed. 
Under more general considerations both quantities are random and described by certain  probability densities \cite{Hughes,Klafter_Sokolov}.
For random walks with independent increments, existence of the second moment of the displacement guarantees that for large number of subsequent steps convergence of the motion to Gaussian diffusion takes place with the mean squared displacement (MSD) growing linearly in time, $\langle r^2(t) \rangle \equiv \langle | \vec{r}(t)-\vec{r}(0) |^2 \rangle \propto t$. Here, $\vec{r}(t)$ denotes the position of a diffusing particle at the moment $t$ and the $\langle \cdot \rangle$ means averaging over independent particles' trajectories. When describing diffusive transport quite commonly the time dependence of the MSD has been used to discriminate between ``normal'' (MSD linear in time) and `anomalous'' (MSD growing sub or superlinearly in time) character of diffusion.  It turns out however \cite{dybiec2009h, sokolov1997, sokolov2000, forte2014} that this kind of qualification, if applied straightforwardly, may be misleading. 
If both, decoupled jump length and waiting times of a one dimensional random walk are distributed with power-law asymptotics, i.e. $p(r)\propto|r|^{-1-\mu}$, $p(t)\propto t^{-1-\nu}$  (with $0< \mu < 2$ and $0<\nu<1$) and $p(r,t)=t^{-\nu / \mu}p(rt^{-\nu / \mu},1)$ it may happen for $2\nu=\mu$ that ensemble average leads to a linear signature of time dependent MSD (paradoxical diffusion), although the underlying process is neither Gaussian nor Markov. In such situations linear in time scaling of MSD  demonstrates competition between long rests and long jumps, leading effectively (for a finite number of steps) to the second moment of the $p(r,t)$ scaling like $t^{2\nu/ \mu}$.

The assumption of the power-law asymptotics of the jump length distribution originates in numerous observations of very distinct systems.
The presence of fluctuations distributed according to heavy-tailed power-law densities have been observed in various situations in
physics, chemistry or biology \cite{shlesinger1995,nielsen2001},
paleoclimatology \cite{ditlevsen1999b,ditlevsen2005} or economics
\cite{mantegna2000}. The heavy-tailed
fluctuations are well visible in various models
\cite{solomon1993,solomon1994,chechkin2002b,boldyrev2003}, and are
studied in an increasing number of situations
\cite{chechkin2004,sokolov2003,dubkov2008,rypdal2010,barthelemy2008,pasternak2009,lomholt2005,klages2008,pasternak2009}.

Two dimensional L\'evy flights are considered as a starting point of optimal random search strategies of sparse, randomly distributed targets, see \cite{Viswanathan1996} and \cite{Edwards2007,gonzalez2008} for a related discussion on applicability and observability of such scenarios. L\'evy flights based search strategies have better statistical properties and consequently result in higher search efficiency than random search strategies based on Gaussian random walks \cite{teuerle2009,teuerle2012}. Trajectories of 2D L\'evy flights have very similar statistical properties to bivariate L\'evy-stable random walks, which generalizes 1D L\'evy-stable random walks, see \cite{samorodnitsky1994}. L\'evy flights are characterized by the unbounded variance. However, presence of targets introduce an effective cut-off to the jump length distribution.

In this paper we investigate properties of the 2D L\'evy flights in crowded confined geometries mimicked by a random mesh of anisotropic molecules randomly placed on a surface. This particular kind of structure used in this study models fibrinogen monolayers on mica surface \cite{adamczyk2010, adamczyk2011} and allows to determine effects of surface blocking functions and jamming coverages \cite{ciesla2013,barbasz2013}. It is also known to model fundamental properties of the protein in formation of blood clots, thrombosis or tumor growth \cite{Mosesson}. Here, however, we will focus on a self-diffusion of a particle meandering within such mesh and performing walk with admissible non-Gaussian (Cauchy) distribution of steps. Our task is therefore to analyze to what extend presence of random obstacles influences the motion \cite{hofling2013, sokolov2012, strzelewicz2013} and slows down the diffusion process. By using a random walk model described in Sec.~\ref{sec:model}, we study role of environmental crowding 
and finite domain of motion on the asymptotic transport properties.

Numerical results and discussion are presented in Secs.~\ref{sec:results} and~\ref{sec:discussion}. The paper is closed with Summary and Conclusions (Sec.~\ref{sec:summary}) making comparative analysis with our former studies \cite{ciesla2014} where standard 2D Wiener-Gaussian process has been used to simulate the motion.
%
%
\section{Model\label{sec:model}}
\begin{figure}[!htb]
\begin{tabular}{c}
\includegraphics[width=0.7\columnwidth]{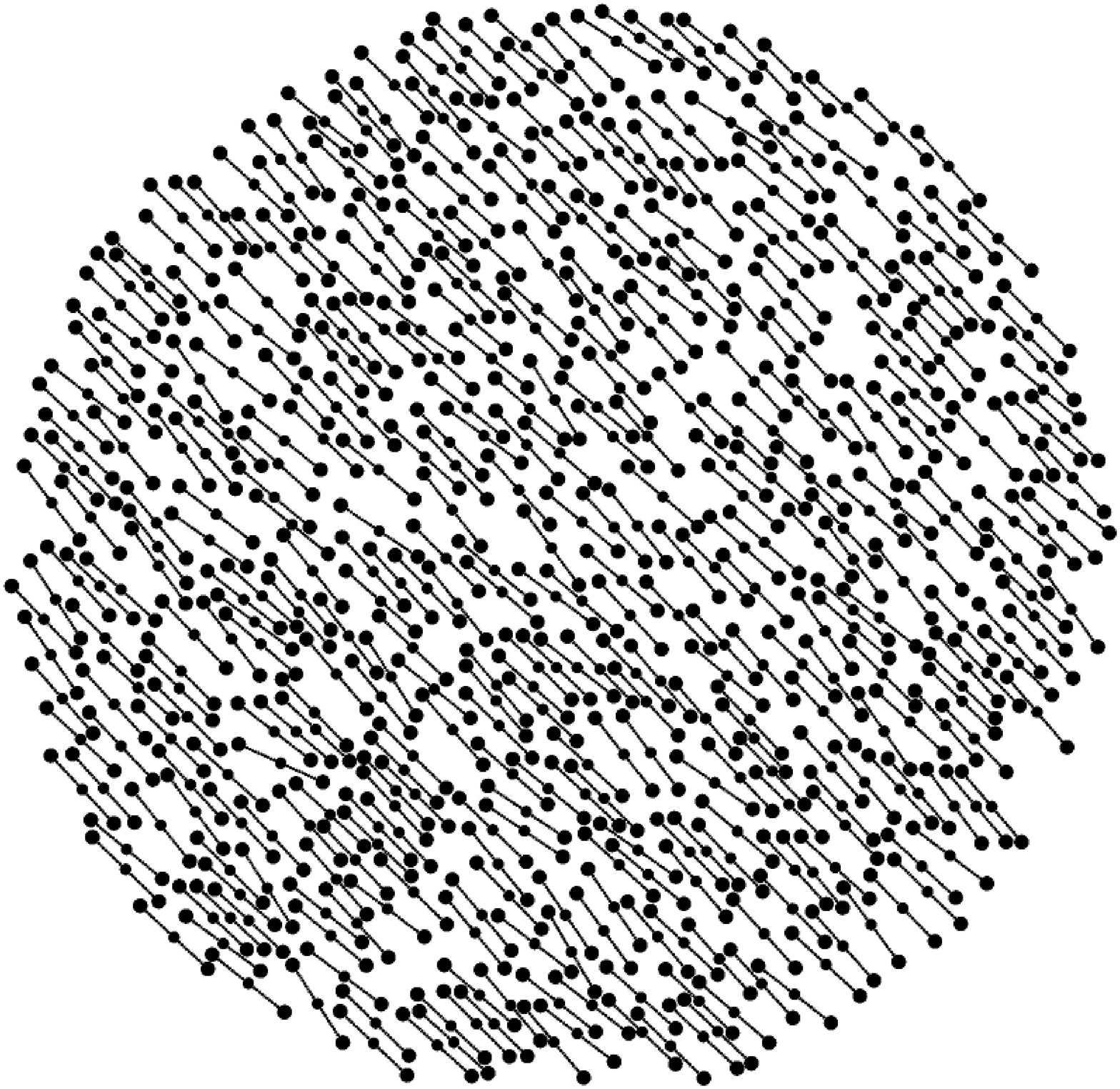}\\
\includegraphics[width=0.7\columnwidth]{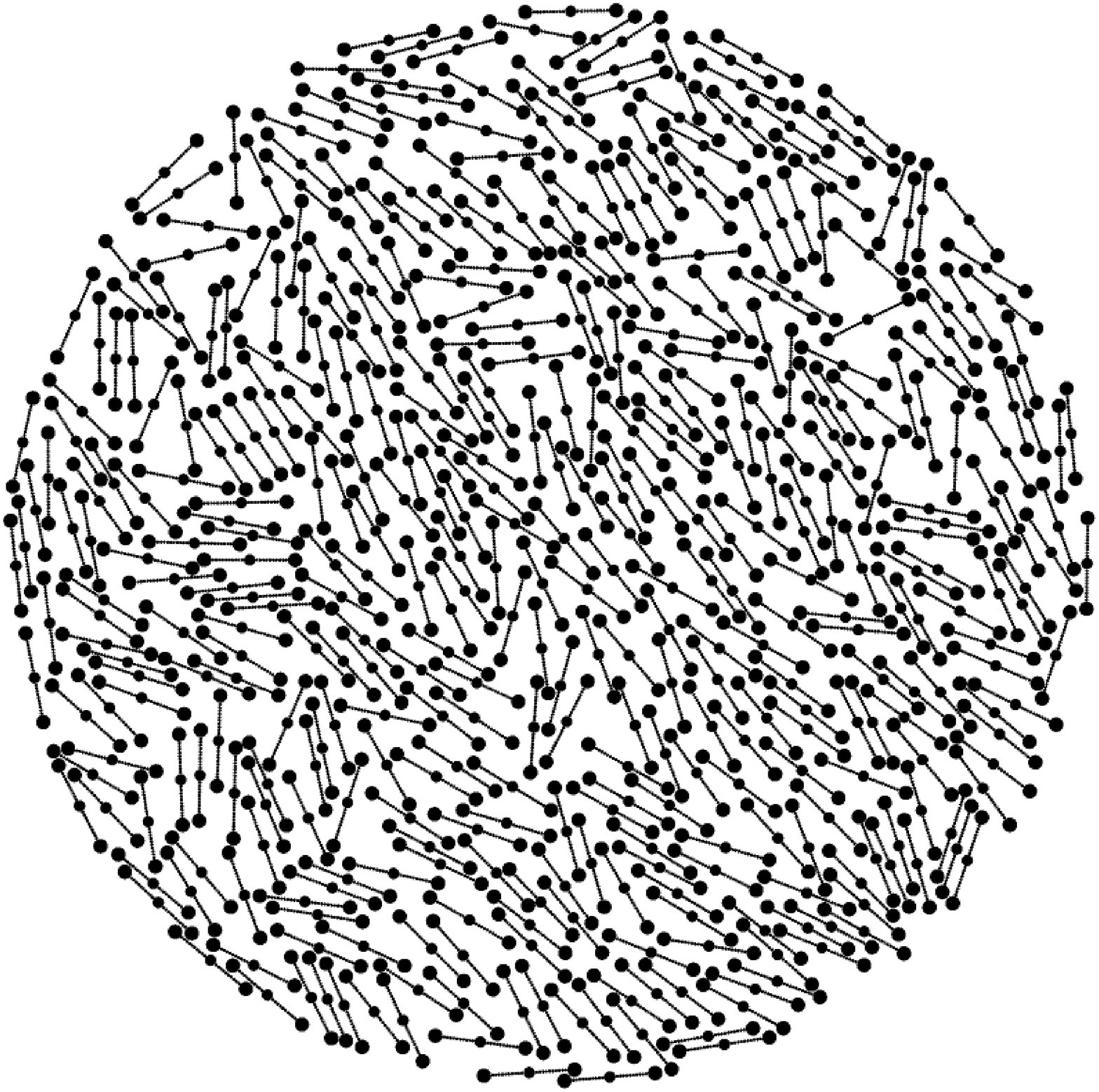}\\
\includegraphics[width=0.7\columnwidth]{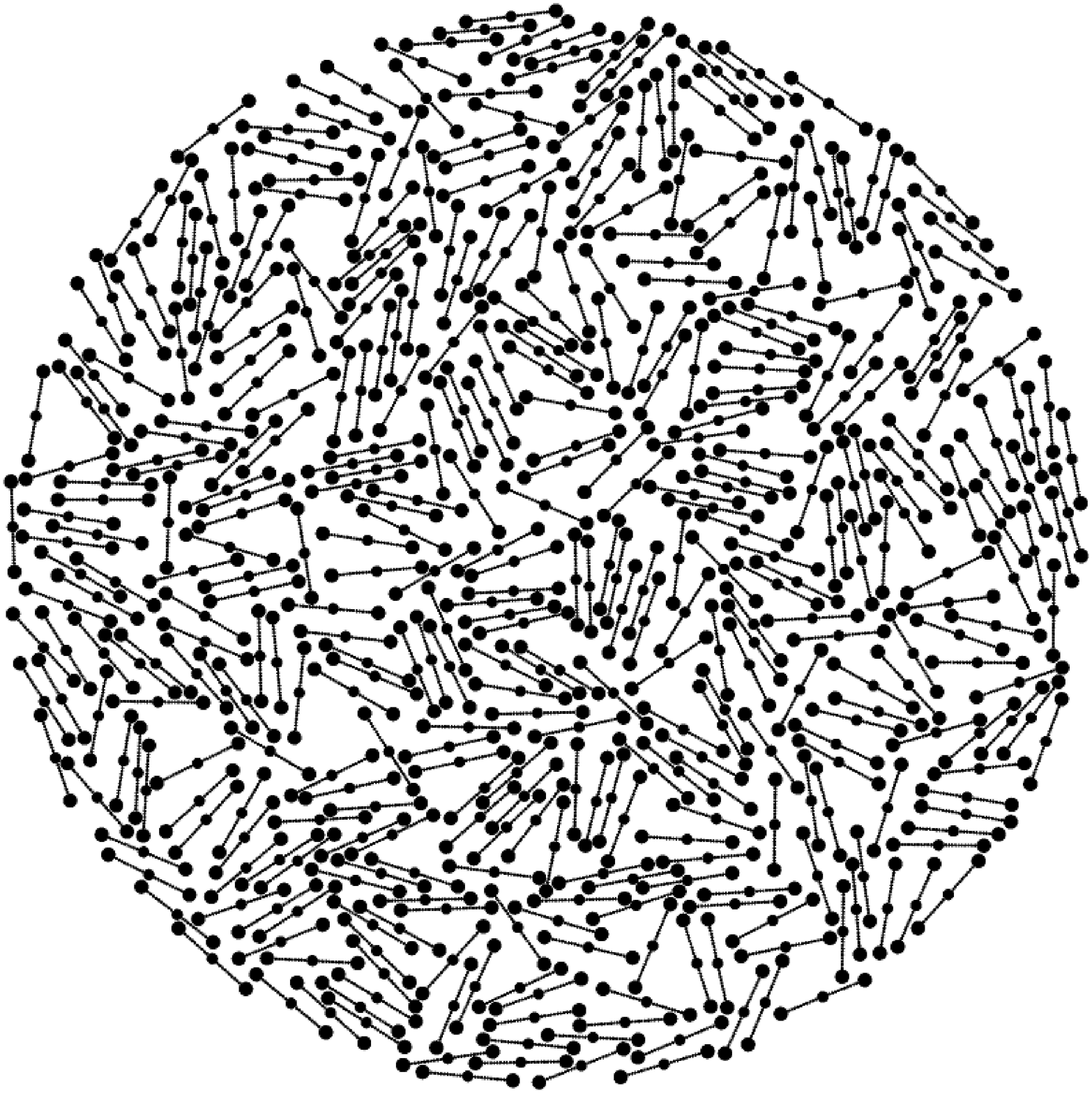}\\
\end{tabular}
\caption{Sample monolayers build of fibrinogen particles characterized by different orientational order: $q=0.98$ (top panel), $q=0.57$ (middle panel), and $q=0.10$ (bottom panel). The diameter of monolayers is $500 \, nm$. The concentration of the fibrinogens is $2140 \, \mu m^{-2}$.}
\label{fig:layers}
\end{figure}

We consider a tracer particle (random walker) whose position $\vec{r}(t)$  is confined in spatially inhomogeneous static cages formed by the mesh of anisotropic obstacles. Transport-related observables are studied by examination of many stochastic trajectories mimicking position of the tracer, constructed by Monte Carlo methods.
As an obstacle we have used here coarse grained approximation of fibrinogen molecule, which help to explain surface density of fibrinogen monolayers obtained as a result of adsorption process \cite{adamczyk2010, adamczyk2011}. Since fibrinogen molecules are strongly anisotropic, they can build orientationally ordered structures whose properties have been analyzed numerically elsewhere \cite{ciesla2013, barbasz2013}.  In brief, the degree of global orientational ordering can be described using the order parameter $q$:
\begin{equation}
q = 2 \left[ \frac{1}{N} \sum_{i=1}^N ( x_i \cos \varphi + y_i \sin \varphi)^2 - \frac{1}{2} \right],
\label{eq:order}
\end{equation}
where $N$ is a number of obstacles $[x_i, y_i]$ is a unit vector parallel to long axis of the $i$-th particle and $\varphi$ is the mean direction of all molecules in a monolayer \cite{ciesla2013}. 
Exemplary monolayers characterized by different values of $q$ are shown in Fig.~\ref{fig:layers}.

The tracer starts its motion at random position near the center of the monolayer. Assuming that the tracer position after $i$-th jump is $\vec{r_i} = [x_i, y_i]$, the next position is calculated according to the following scheme:
\begin{itemize}
\item[--] a random number $\xi$ is drawn according to the Cauchy distribution with the scale parameter $\gamma$
\begin{equation}
p \left( x \right) = \frac{\gamma}{\pi \left[ \gamma^2 + x^2 \right]},
\label{eq:cauchypdf}
\end{equation}
\item[--] a direction $\varphi$ is generated according to the uniform distribution $U(0,2\pi)$,
\item[--] the displacement $\Delta \vec{r}$ is calculated as $\Delta\vec{ r} = [\xi \cos \varphi, \xi \sin \varphi ]$,
\item[--] if the straight line from $\vec{r}_i$ to $\vec{r}_i + \Delta\vec{ r}$ does not intersect any obstacle and does not end on an obstacle or outside the system the move is accepted: $\vec{r}_{i+1} = \vec{r}_i + \Delta\vec{ r}$,
otherwise the tracer does not move, i.e. $\vec{r}_{i+1} = \vec{r}_i$.
\end{itemize}
Accordingly, in the obstacle free case, the probability distribution function of the displacement $\Delta \vec{r}$ is spherically symmetric $p(\Delta \vec{r})=p(|\Delta \vec{r}|)$ and after $N$ steps, the position of the walker is given by $\vec{r}_N\propto\sum_{i=1}^N\Delta \vec{r}_i\propto N\propto t$, such that the characteristic function $\langle \exp(ik \vec{r}) \rangle=\hat{p}_{\vec{r}(k, 1, \gamma)}=\exp(-\gamma|\vec{k}|)$. Hence L\'evy (Cauchy) flights performed by the walker disperse faster than $N^{1/2}$.

For each orientational order, $q$, of obstacles $20$ to $100$ images of different fibrinogen layers were analyzed. For each one of monolayer setups $3000$ independent paths consisting of $10^4$ jumps were generated. As the tracers move on a two dimensional images the natural unit of a distance used is one pixel (here it corresponds to $0.7 \, nm$), which is also the tracer diameter. Therefore, if not explicitly specified, all distances in this study will be denoted in pixels. Similarly, the single iteration of the above path generation procedure takes the role of the time unit, i.e. time is measured in the number of jumps.
Parameter $\gamma$ in Eq.~(\ref{eq:cauchypdf}) has been set to $0.1$.

\begin{figure}[!htb]
\vspace{1cm}
\centerline{%
\includegraphics[width=0.8\columnwidth]{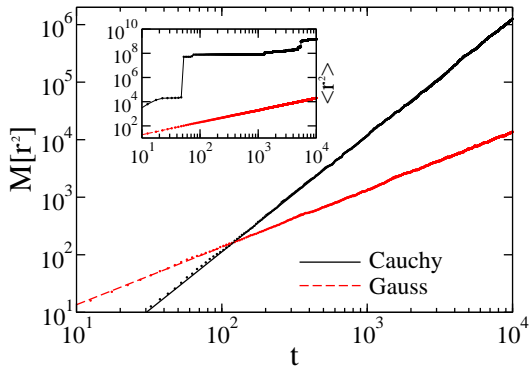}}
\caption{(Color online) Median of the squared displacement for 3000 independent tracers. Black points correspond to displacements generated by Cauchy distribution~(\ref{eq:cauchypdf}), whereas red points are obtained using Gaussian normal distribution $N(0, 1)$. Lines represents fits: $M[r^2]=0.01 \cdot t^{2.02}$ and $M[r^2]=1.35 \cdot t^{1.00}$ for Cauchy and Gauss distributions, respectively. Inset shows the mean squared displacement for both types of diffusion.}
\label{fig:test_free}
\end{figure}

To check if the above procedure generates superdiffusive process we test it using first unrestricted empty space available to the randomly walking tracer. Results presented in Fig.~\ref{fig:test_free} confirm that Cauchy distributed jumps indeed result in the superlinear growth of the MSD. 
The apparent disruption of the sample MSD is a direct consequence of the fact that, contrary to the Gaussian case, trajectories generated using Cauchy distribution are discontinuous. In order to avoid the abrupt, discontinuous changes of the MSD one can use as a measure of the tracer displacement the the median $M[r^2]$:  In general,  quantiles $Q_m$ ($0<m<1$) are defined according to the relation $F(Q_m,t)=m$, where
\begin{equation}
F(x,t)=\int^{x}_{-\infty}p(z',t)dz'
\end{equation}
is the cumulative distribution function (CDF) of a random variable $x$. By choosing $m=1/2$ and  $x\equiv r^2$, the median of the square displacement fulfills
\begin{eqnarray}
\mathrm{Prob}\left\{r^2(t)\leq M[r^2]\equiv Q_{0.5}\right\}=\frac{1}{2}.
\end{eqnarray}

In the case under investigation (cf. Fig.~\ref{fig:test_free}), the median of the squared displacement grows like $t^{2.02}$ thus indicating superdiffusive character of Cauchy flights. At the same time, if jump lengths are distributed according to the normal density, the median grows linearly in time demonstrating normal diffusive character of the 2D Wiener process.

It is worth to emphasize that the Cauchy distribution is characterized by the diverging variance and its mean value can be defined only in a generalized sense as a ``principal value'' of the integral. Therefore, the sample MSD depends on a specific set of trajectories and can display discontinuities (see inset of Fig.~\ref{fig:test_free}), in contrast to the median $M[r^2]$ presented in the main plot.

%
%
\section{Results\label{sec:results}}
\begin{figure}[!htb]
\centerline{%
\includegraphics[width=0.45\columnwidth]{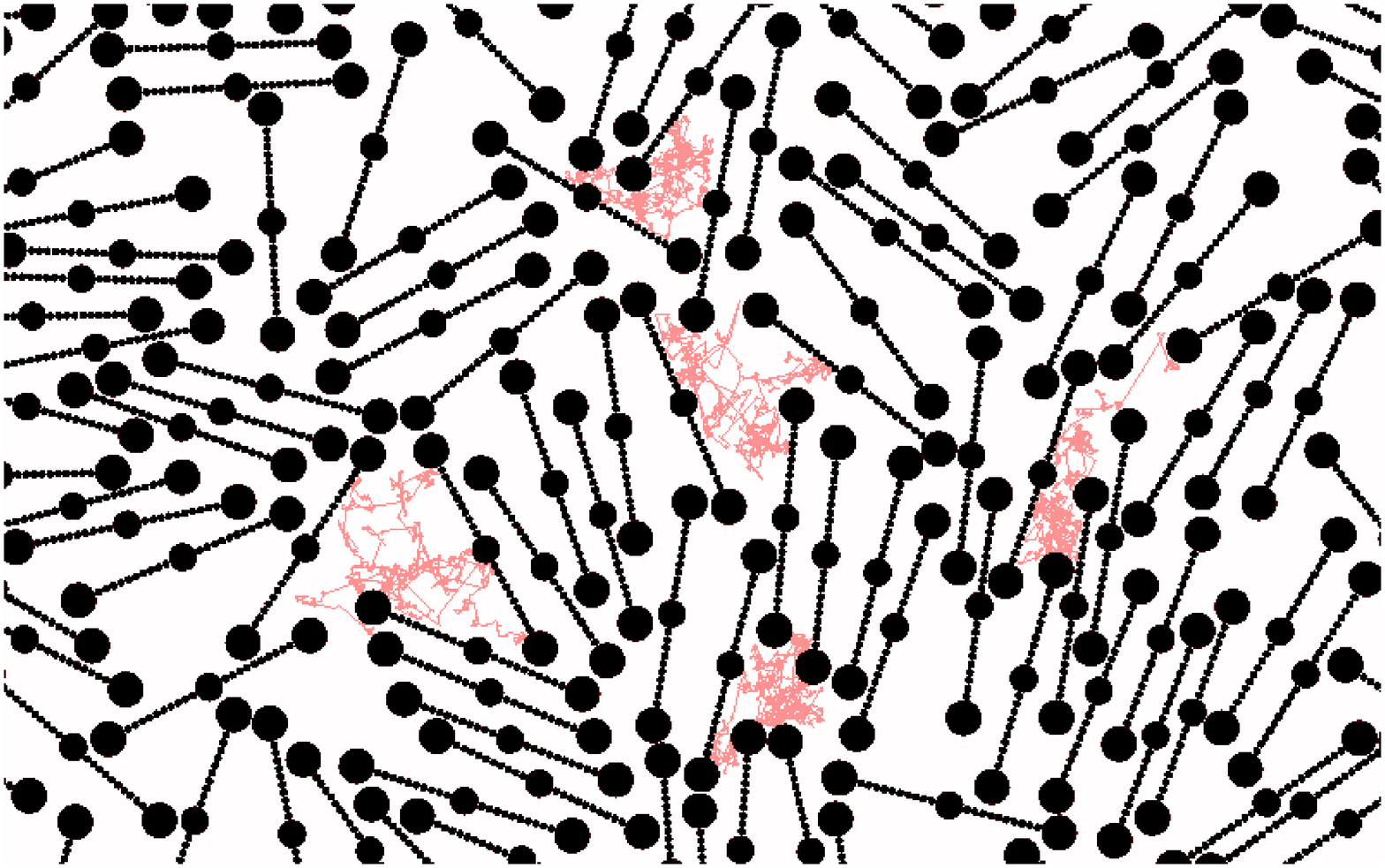}
\hspace{0.3cm}
\includegraphics[width=0.45\columnwidth]{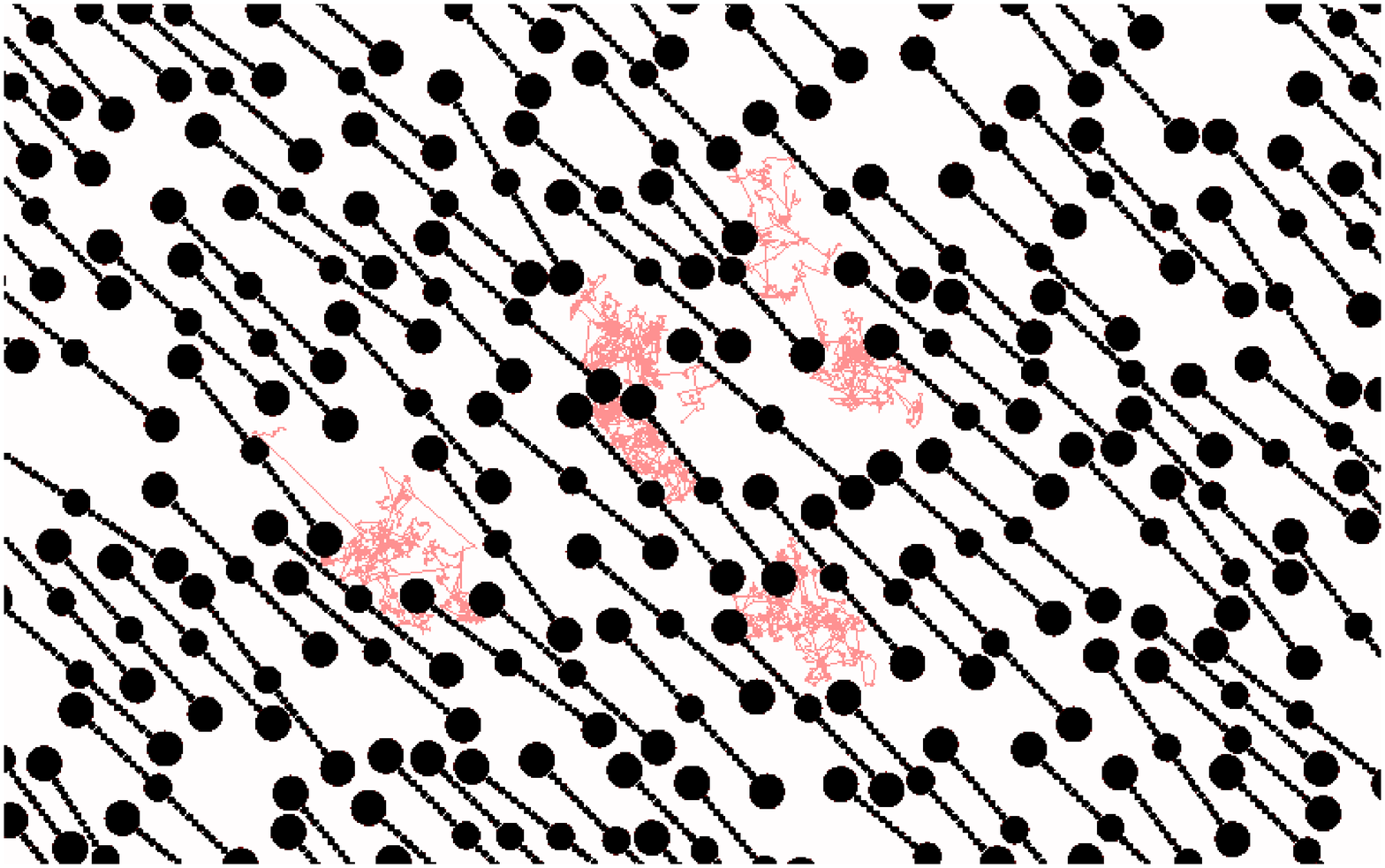}
}
\vspace{0.5cm}
\centerline{%
\includegraphics[width=0.45\columnwidth]{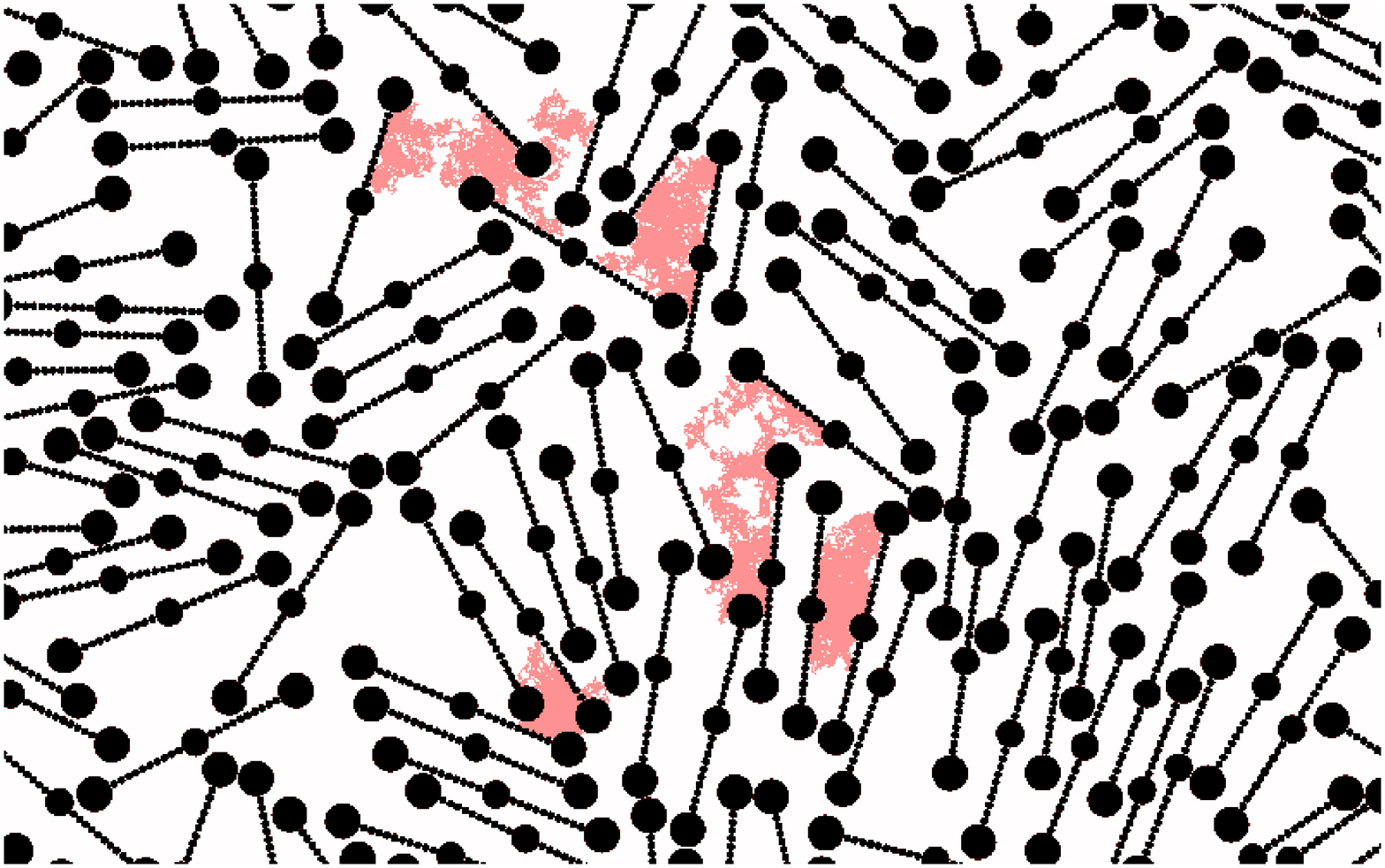}
\hspace{0.3cm}
\includegraphics[width=0.45\columnwidth]{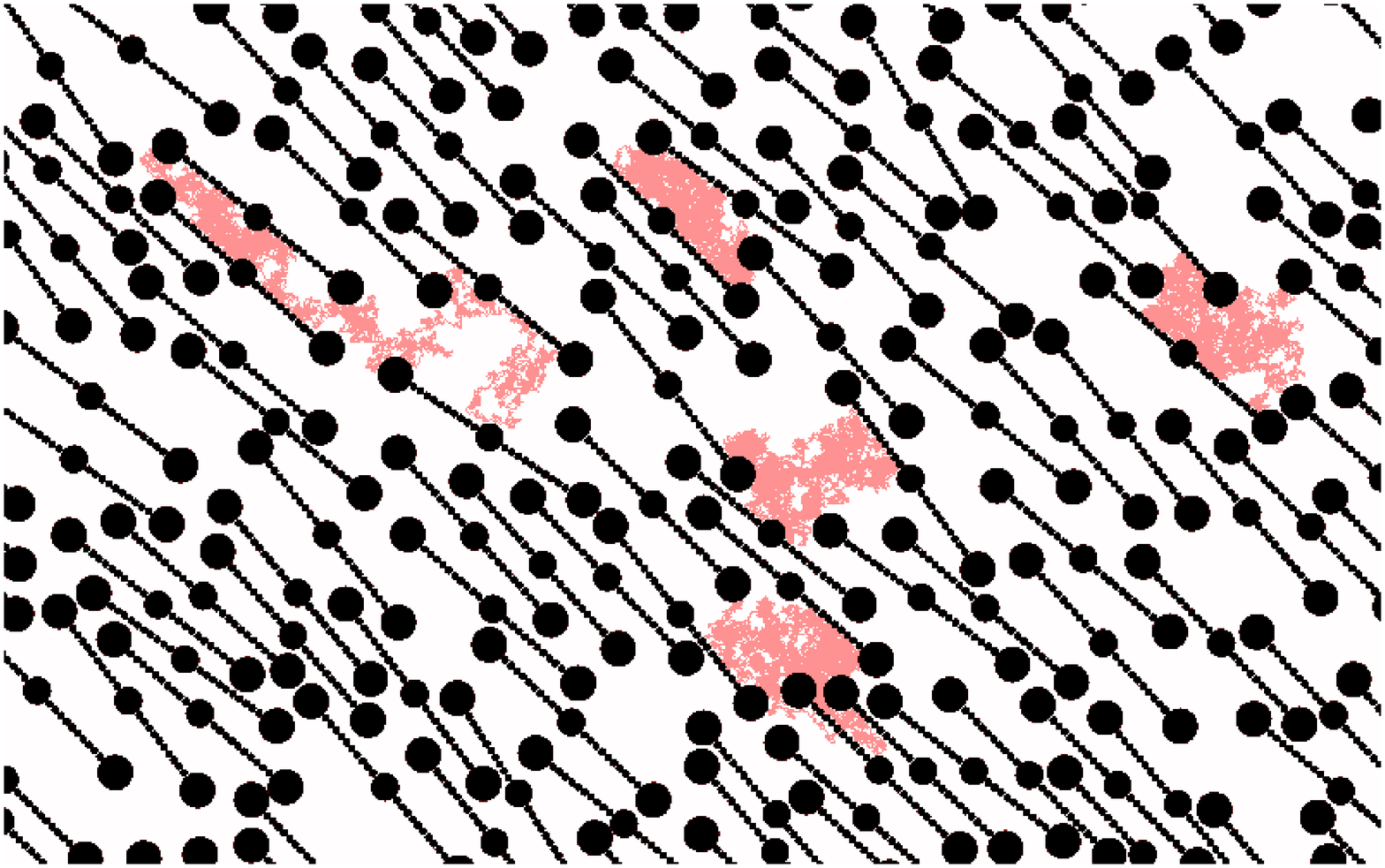}
}
\caption{(Color online) Sample paths of five independent tracers inside orientationally unoriented ($q=0.1$, left) and oriented ($q=0.98$, right) fibrinogen mesh. Trajectories on the upper pictures were generated by using Cauchy distribution (\ref{eq:cauchypdf}). Bottom panels show similar trajectories produced by Gaussian random walk. Each trajectory contains $10^4$ steps. The concentration of the fibrinogens is $2140 \, \mu m^{-2}$. }
\label{fig:trajectories}
\end{figure}
Figure \ref{fig:trajectories} depicts exemplary 2D trajectories for a particle motion within a fibrinogen monolayer with Gauss and Cauchy distributions of step lengths. Typically for the L\'evy processes 
\begin{equation}
\lim_{\Delta t\rightarrow 0} \mathrm{Prob}\left\{|x_i(t+\Delta t)-x_i(t)|>\epsilon\right \}=0 
;\;\;\forall \epsilon>0
\end{equation}
leading to local clustering of trajectories interspersed with occasional long jumps.

\begin{figure}[!htb]
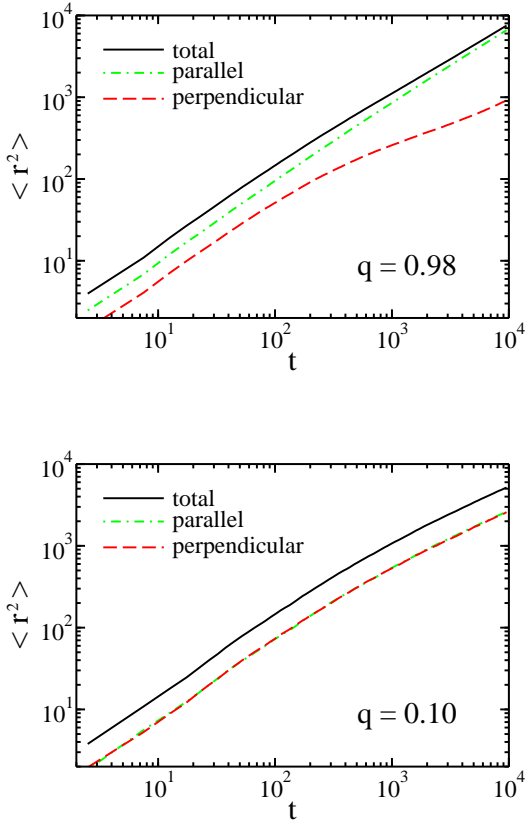

\vspace{1cm}
\centerline{
\includegraphics[width=0.8\columnwidth]{msd_t_ordered}}
\vspace{1cm}
\centerline{
\includegraphics[width=0.8\columnwidth]{msd_t_disordered}}
\caption{(Color online) The mean squared displacement and its parallel and perpendicular components in an ordered ($q=0.98$, top panel) and disordered ($q=0.1$, bottom panel) environment. The concentration of obstacles is $2140 \, \mu m^{-2}$.}
\label{fig:msd_t_order}
\end{figure}

\begin{figure}[!htb]
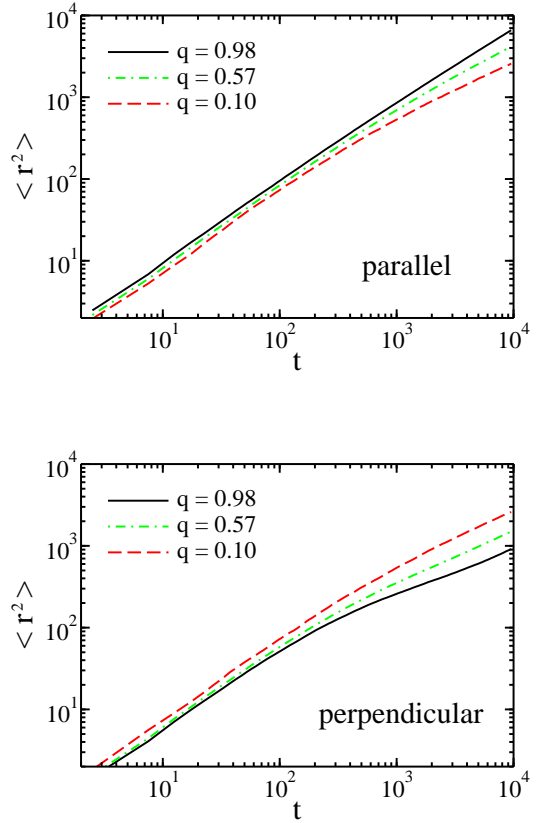

\vspace{1cm}
\centerline{
\includegraphics[width=0.8\columnwidth]{msd_t_parallel}}
\vspace{1cm}
\centerline{
\includegraphics[width=0.8\columnwidth]{msd_t_perpendicular}}
\caption{(Color online) Parallel (top panel) and perpendicular (bottom panel) components of the mean squared displacement for different orientational order of obstacles $q$. The concentration of obstacles is $2140 \, \mu m^{-2}$. }
\label{fig:msd_t_parper}
\end{figure}

\begin{figure}[!htb]
\vspace{1cm}
\centerline{
\includegraphics[width=0.8\columnwidth]{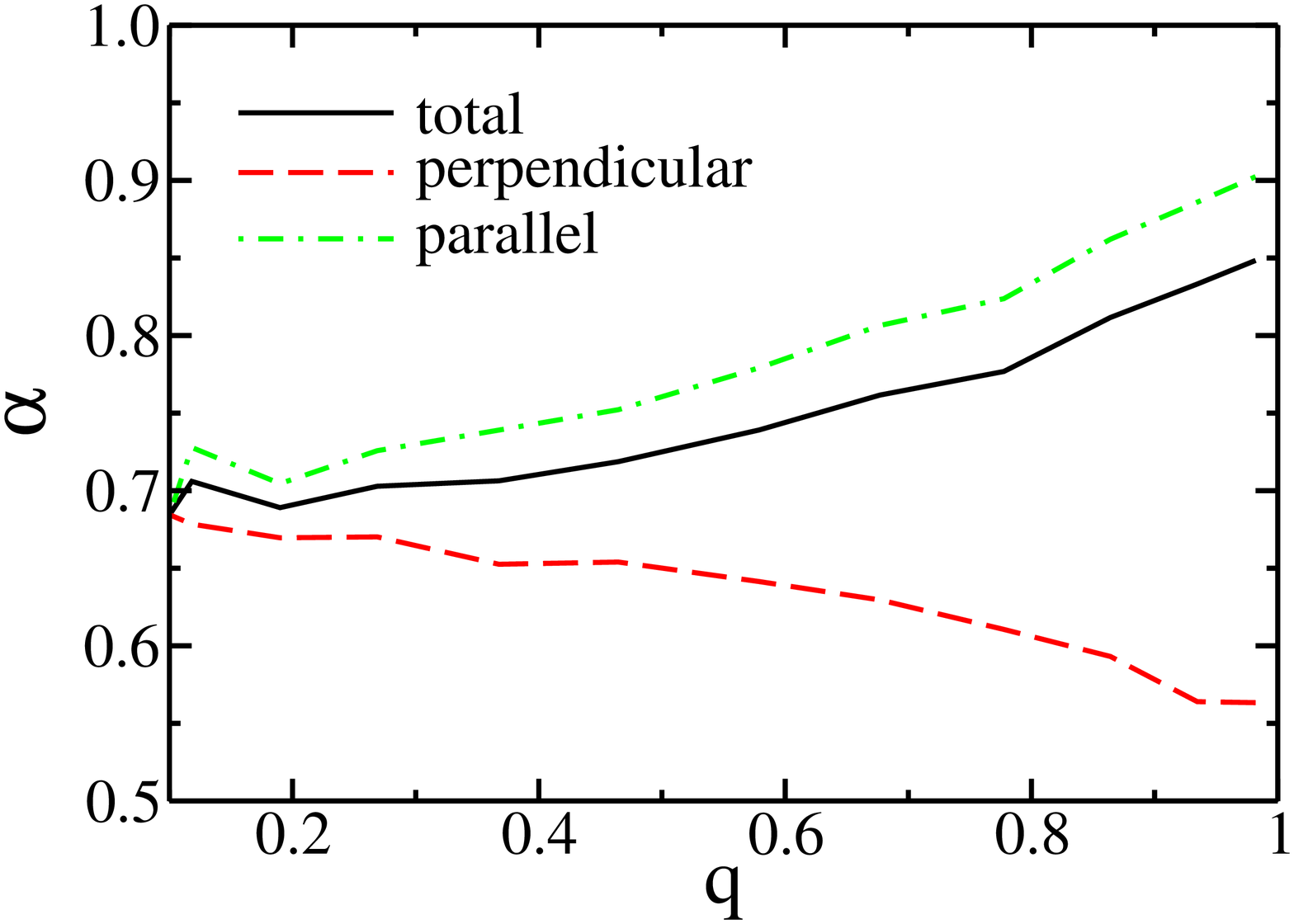}}
\vspace{1cm}
\centerline{
\includegraphics[width=0.8\columnwidth]{a_ordered}}
\caption{(Color online) Top panel presents the effective exponent $\alpha$ as a function of the obstacle orientational order parameter $q$ for total diffusion as well as for its parallel and perpendicular components. The value of $\alpha$ is averaged over $t \in \{10, 10^4\}$. The bottom panel demonstrates the dependence of the effective exponent $\alpha$ on time for highly orientationally ordered obstacles. The concentration of obstacles is $2140 \, \mu m^{-2}$.}
\label{fig:msd_t_ordered}
\end{figure}

The free space formed by obstacles, in which the random walk  takes place, can be highly anisotropic. In consequence, the MSD can be expressed in the form of two components, which measure diffusion along the direction of the average obstacles orientation and in the direction perpendicular to it:
\begin{equation}
\langle r^2(t) \rangle = \langle r^2(t)\rangle_{\parallel} + \langle r^2(t)\rangle_{\perp}.
\label{eq:components}
\end{equation}
Environment anisotropy is produced by different ordering of fibrinogen monolayers, see Eq.~(\ref{eq:order}).
Numerical results, corresponding to various level of obstacles' orientational order (see Fig.~\ref{fig:layers}), are shown in Figs.~\ref{fig:msd_t_order}, \ref{fig:msd_t_parper} and~\ref{fig:msd_t_ordered}.

Figure~\ref{fig:msd_t_order} presents dependence of the MSD and its components (parallel and perpendicular) for high (top panel) and low (bottom panel) values of the order parameter $q$. If the order parameter is low, there is no preferred direction in the system and consequently perpendicular and parallel components of the MSD are indistinguishable. With increasing order, both components of the MSD start to differ. Clear acceleration of the motion along fibrinogen direction is observed, what is confirmed by larger values of the MSD parallel to average molecule directions in comparison to the perpendicular one.
This effect is further corroborated in Fig.~\ref{fig:msd_t_parper} which displays dependence of parallel (top panel) and perpendicular (bottom panel) components of the MSD for various values of the order parameter $q$. With increasing value of the order parameter $q$ the parallel component of the MSD increases whereas its perpendicular part decreases. This behavior is a signature of increasing ordering of elongated molecules forming a monolayer.
For low $q$ most molecules are randomly oriented, while for large $q$ they form an aligned structure of channels favoring motion along fibrinogen molecules.

Asymptotically (for large number of steps $N$ or equivalently, at sufficiently long times $t\rightarrow\infty$), the passive random walk describing position of the tracer $\vec{r}_N\propto\sum_{i=1}^N\Delta \vec{r}_i$ tends to a stochastic diffusion process. When obstacle mesh forms orientationally ordered structure, this diffusion becomes strongly anisotropic with hindered perpendicular dislocations. In order to further analyze this pattern, the measured MSD dependence on time has been compared with a power-law
\begin{equation}
\langle r^2(t) \rangle = D t^\alpha,
\label{MSD_ALPHA}
\end{equation}
where $D$ is a diffusion coefficient, and the exponent $\alpha$ describes diffusion type.
Commonly, the exponent $\alpha$ has served to discriminate between normal ($\alpha=1$) and anomalous diffusion (with $\alpha<1$ corresponding to subdiffusion and $\alpha>1$ to superdiffusion, respectively). 
However, as mentioned already in the Introduction, sole use of formula (\ref{MSD_ALPHA}) does not qualify the process unambiguously.
There are anomalous transport motions \cite{dybiec2009h, sokolov1997, sokolov2000, forte2014}, in which  competition between long trapping events and long jumps might result in sample MSD increasing linearly in time, thus hiding the subdiffusive character of motion. Therefore, the parameter $\alpha$ should be treated as an effective diffusion exponent

The effective exponent $\alpha$ can be also determined using the logarithmic derivative:
\begin{equation}
\alpha = \frac{d \ln \langle r^2(t) \rangle}{d \ln t},
\end{equation}
which allows to determine temporal values of $\alpha$. The results of effective exponent $\alpha$ analysis for the diffusion between orientationally ordered obstacles is shown in Fig.~\ref{fig:msd_t_ordered}.

Top panel of Fig.~\ref{fig:msd_t_ordered} clearly indicates that with increasing order parameter $q$ the major component contributing to the effective exponent $\alpha$ originates from the parallel part of the MSD. For ordered phases ($q=0.98$) the perpendicular effective exponent $\alpha$ displays a broad minimum for intermediate times (cf. bottom panel of Fig.~\ref{fig:msd_t_ordered}) which stays in line with observation of the plateau in the perpendicular component of the MSD (see bottom panel of Fig.~\ref{fig:msd_t_parper}) and signalizes  confinement of the tracer in long channels formed by obstacles.  Note, that for relatively short times, $\alpha \approx 1$, indicating that initially tracers do not experience molecular crowding \cite{sokolov2012} which obscures the motion only at longer times. 


%
%
\section{Discussion\label{sec:discussion}}

Values of the effective exponent $\alpha<1$ characterizing the power-law scaling of the MSD suggest that the diffusion process in a crowded environment is subdiffusive despite the fact that displacements are drawn from the Cauchy distribution allowing higher occurrence of extreme jumps than in the case of common Gaussian statistics. Quantitatively the similar results have been reported in our former studies \cite{ciesla2014}, where the tracer movements were sampled from two dimensional Gaussian distribution. In particular, the dependence of MSD on orientational order of obstacles and their concentration are strikingly alike in both cases. This observation suggests that at the level of the MSD analysis, the diffusion type in crowded environment is determined mostly by local disorder. Moreover,  lack of significant differences in obtained results between realizations of Gauss or Cauchy  random walks is due to truncation of allowed jumps by obstacles: Within the proposed random walk model 
it has been assumed that a tracer cannot neither cross an obstacle or escape from the system. In other words, jumps which cross fibrinogen molecule or end outside the system have been rejected from the statistical analysis. In consequence, such a procedure  leads to truncation of tails in the underlying Cauchy distribution of step lengths and homogenizes the random walk model.

\begin{figure}[!htb]
\vspace{1cm}
\centerline{
\includegraphics[width=0.8\columnwidth]{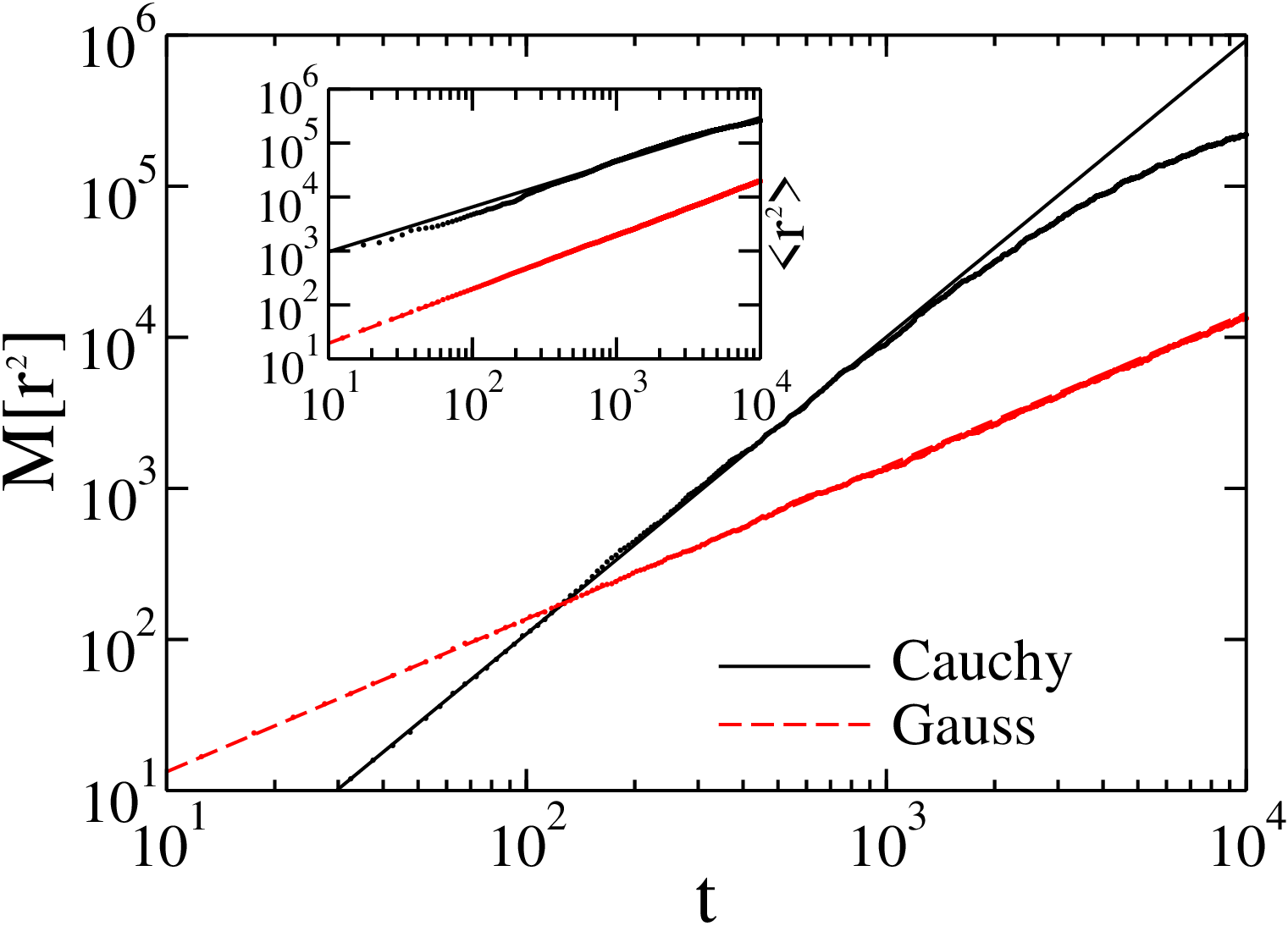}}
\vspace{1cm}
\centerline{
\includegraphics[width=0.8\columnwidth]{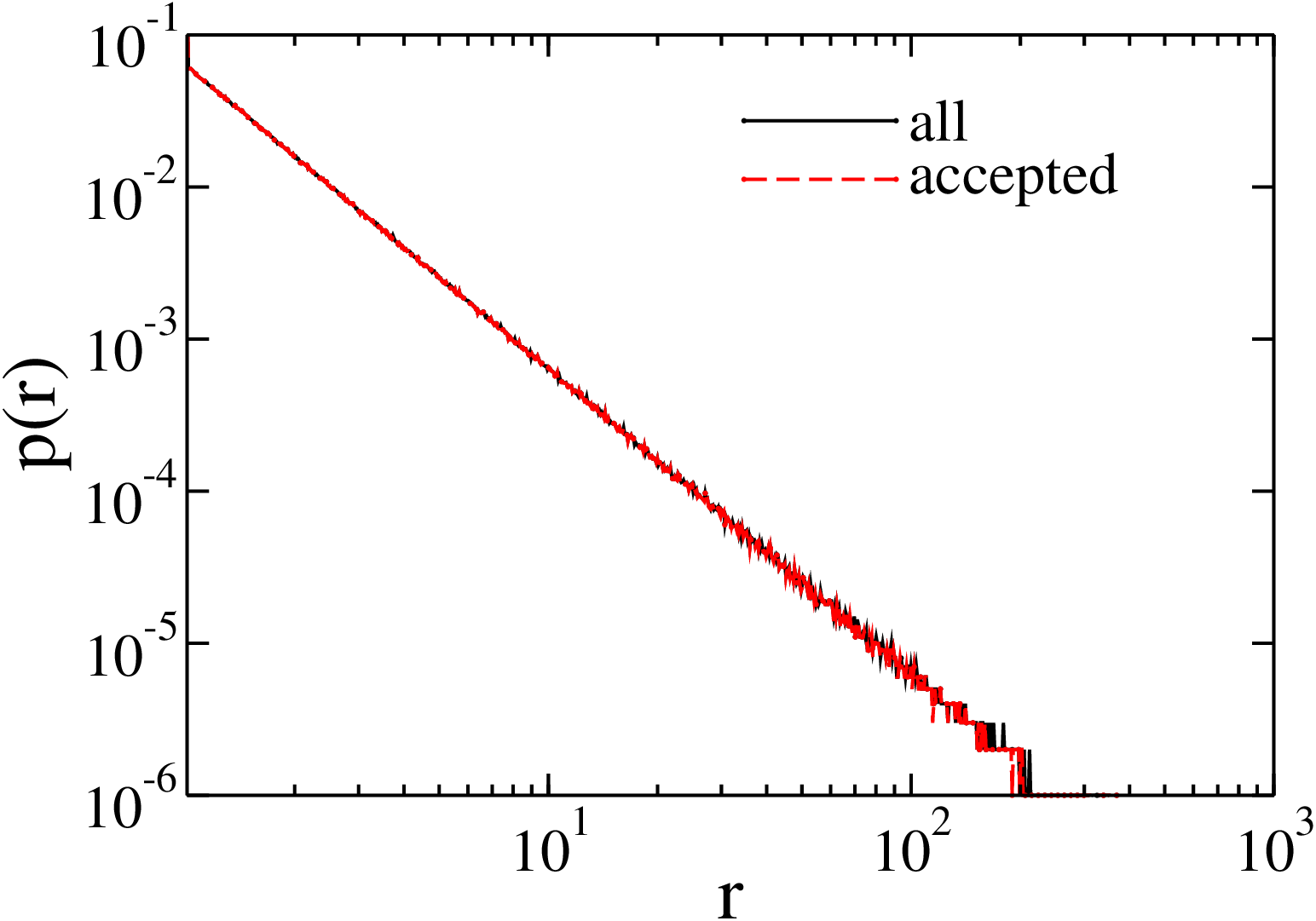}}
\caption{(Color online) Top plot shows median of the squared displacement for 3000 independent tracers in a restricted empty space. Black points corresponds to displacements generated using Cauchy distribution~(\ref{eq:cauchypdf}) whereas red points are obtained using Gaussian distribution $N(0, 1)$. Lines represents fits: $M[r^2]=0.01\cdot t^{1.96}$ and $M[r^2]=1.30 \cdot t^{1.01}$ for Cauchy and Gauss distributions respectively. Inset shows the mean squared displacement of for both types of diffusion. The bottom panel shows histogram of all generated trial displacements according to the Cauchy distribution~(\ref{eq:cauchypdf}) (black solid line) and accepted ones (red dashed line).}
\label{fig:test_limited}
\end{figure}

In order to further discriminate between effects of the finiteness of the system and presence of obstacles, we have first  analyzed properties of free diffusion in restricted space, see Fig.~\ref{fig:test_limited}. 
Top panel of Fig.~\ref{fig:test_limited} presents dependence of the median of the squared displacement in a confined empty geometry.
For short and moderate times, i.e. up to $t\approx 10^3$, independent Cauchy tracers show signature of group superdiffusion, i.e. $M[r^2(t)] \propto t^\alpha$ with $\alpha \approx 2$.
For longer times,  boundary effects become important and the group velocity $dM[r^2]/dt$  drops down.
In contrast, the Gaussian jump length distribution leads to slower diffusion for which median $M[r^2]$ grows linearly in time and within the analyzed time window does not exhibit saturation induced by boundary effect.
Although median dependence on time indicates that both types of diffusion processes are significantly different, such a discrimination between normal and anomalous motion is impossible when studying solely the MSD time dependence: For both Gaussian and Cauchy random walks slopes of MSD represented in the log-log scale remain similar, see inset in Fig.~\ref{fig:test_limited}. Moreover, within the time window (up to $t=10^4$) almost all jumps of the Cauchy random walk are accepted. This is confirmed by histograms of generated and accepted movements, which are depicted in the bottom panel of Fig.~\ref{fig:test_limited}. The truncation due to finite domain of motion is hardly observed. The acceptance ratio is over $0.999$, i.e. the rejection rate is smaller than $10^{-3}$. This indicates that, within a studied time window, a random walker practically does not feel the boundaries.

\begin{figure}[!htb]
\vspace{1cm}
\centerline{
\includegraphics[width=0.8\columnwidth]{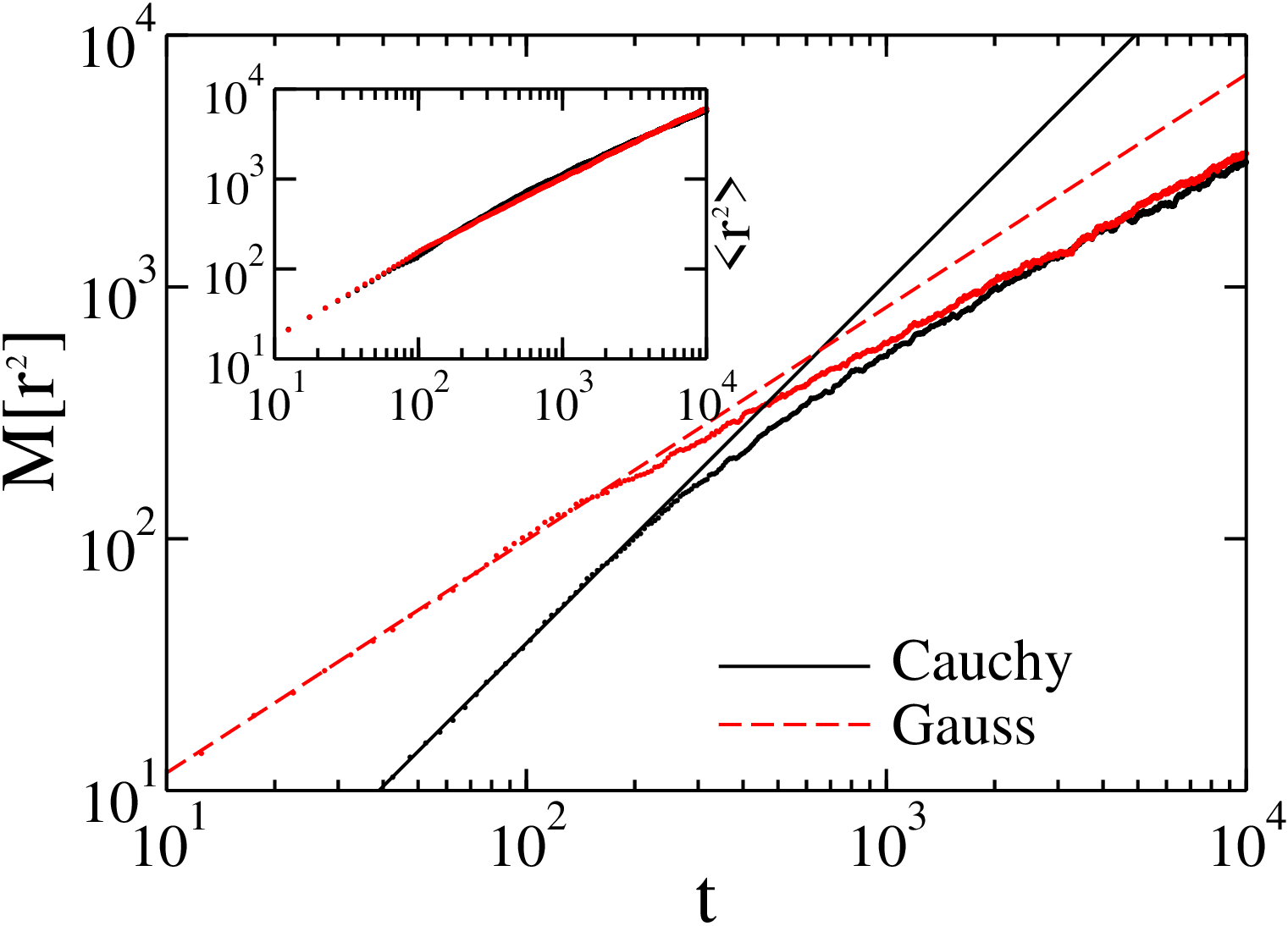}}
\vspace{1cm}
\centerline{
\includegraphics[width=0.8\columnwidth]{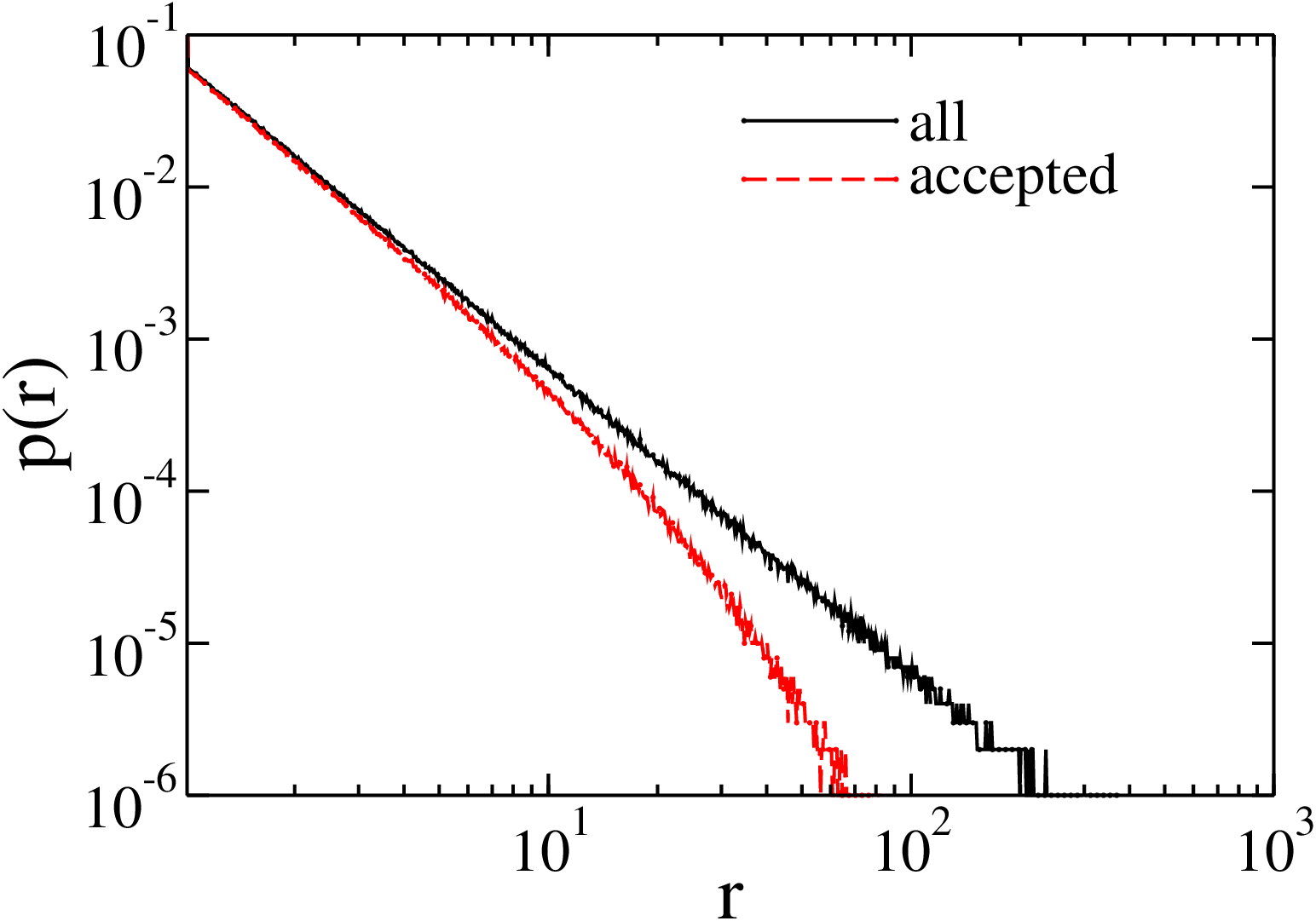}}
\caption{(Color online) Top plot shows median of the squared displacement for 3000 independent tracers in a restricted space filled with unoriented fibrinogen molecules. Black points corresponds to displacements generated using Cauchy distribution~(\ref{eq:cauchypdf}) whereas red points are obtained using Gaussian distribution $N(0, 1)$. Lines represents fits: $M[r^2] = 0.05\cdot t^{1.43}$ and $M[r^2]=1.40 \cdot t^{0.92}$ for Cauchy and Gauss distributions respectively. Inset shows the mean squared displacements for both types of diffusion. The bottom panel shows histogram of all generated trial displacements according to the Cauchy distribution~(\ref{eq:cauchypdf}) (black solid line) and accepted ones (red dashed line). The concentration of obstacles is $2140 \, \mu m^{-2}$.}
\label{fig:test_crowded}
\end{figure}

Figure~\ref{fig:test_crowded} compares 2D Cauchy flights with 2D Wiener process in confined, crowded geometry.
Initially a significant difference between scaling of medians of squared displacements is visible, however, due to large amount of obstacles the diffusion becomes hindered. For longer times the median does not provide distinction between Gaussian random walks and L\'evy flights, whereas the MSD remains  the same for all times.
Similar  asymptotic scaling of median for both type of motion is a consequence of accumulating number of tracer's contacts with the domains of fibrinogen molecules.
The truncation effects turn to be so strong that the detailed type of the jump length distribution is not longer important.
The main role is played by surrounding obstacles which effectively  limit maximal jump range.

\begin{figure}[!htb]
\vspace{1cm}
\centerline{%
\includegraphics[width=0.8\columnwidth]{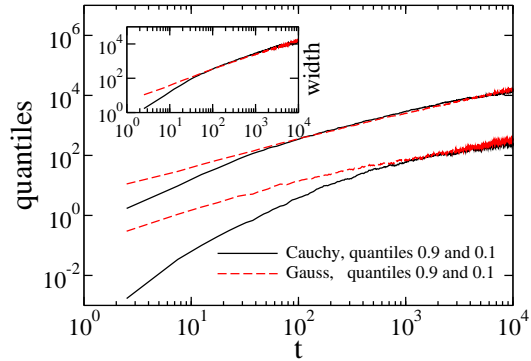}
}
\caption{(Color online) Quantiles $Q_{0.1}(t)$ and $Q_{0.9}(t)$ of the squared displacement for 3000 independent tracers in a limited space full of unoriented fibrinogens molecules. Solid black lines corresponds to displacements generated using the Cauchy distribution~(\ref{eq:cauchypdf}) whereas dashed red lines are obtained using the Gaussian distribution $N(0, 1)$. Inset shows the interquantile width of the squared displacement distribution defined as the difference between quantiles $Q_{0.9}(t)$ and $Q_{0.1}(t)$. The concentration of obstacles is $2140 \, \mu m^{-2}$.}
\label{fig:test_crowded_quantiles}
\end{figure}

As a result, the system shifts to the domain of attraction of the standard central limit theorem yielding unified diffusive behavior for various jump length distributions.
Histograms of generated and accepted movements in L\'evy (Cauchy) flight cases show clearly that longer jumps are rejected, although the number of rejected displacements is quite small, i.e. it does not exceed 2\%, see bottom panel of Fig.~\ref{fig:test_crowded}. Similar conclusions can be drawn from examination of other quantiles of the tracer squared displacement. 
Fig.~\ref{fig:test_crowded_quantiles} displays quantiles $Q_{0.1}(t)$ and $Q_{0.9}(t)$ (main plot) and  the interquantile width $\Delta Q(t)\equiv Q_{0.9}(t)-Q_{0.1}(t)$ (see the inset). 
The behavior of quantiles indicates that for small and intermediate times both processes differ significantly and become indistinguishable after transient period at $t>10^3$, when obstacles hinder duration of trajectories of Gauss and Cauchy jumps.

Taken from the perspective of the continuous time random walk (CTRW), the probability of finding the tracer particle at time $t$ at position $\vec{r}$ can be expressed in terms of probability $\phi(t)$  to stay in the reached position at time $t$ and probability $\eta(\vec{r},t)$ to find a tracer at $\vec{r}$ immediately after the step has been taken \cite{Klafter_Sokolov}
\begin{equation}
p(\vec{r},t)=\int_0^td\tau\phi(t-\tau)\eta(\vec{r},\tau).
\end{equation}
Probability $\phi(t)$ is related to the probability density $\psi(t)$ to wait time interval $t$ between two consequent steps:
\begin{equation}
\phi(t)=\int _t^{\infty}ds\psi(s)
\end{equation}
and for $\psi_N(t)$ being the probability density to make the $N$th jump at time instant $t$ leads to the following recursive relation
\begin{equation}
\psi_{N+1}(t)=\int_0^t d\tau \psi(t-\tau)\psi_N(\tau)\;\;\;\;\psi_1(t)\equiv \psi(t),
\end{equation}
so that $\eta(\vec{r},t)=\sum_{N=1}^{\infty}\psi_N(t)p_N(\vec{r})$ where $p_N(\vec{r})$  stands for probability density to find a particle at position $\vec{r}$ after executing $N$ steps.

Typically  in this scenario subdiffusion  is associated with slowly decaying memory, i.e. the waiting time distribution $\psi(t)$ is expected to have heavier asymptotics than exponential decay, $\psi(t)\sim\frac{\alpha \tau^{\alpha}}{\Gamma(1-\alpha)t^{1+
\alpha}}$ which results in diverging the mean waiting time for $\alpha<1$ and the MSD scaling $\langle r^2 \rangle\sim t^{\alpha}$.
This long memory introduces interdependence (correlation) of increments in the CTRW motion. Therefore, in order to further discriminate between non-Markov subdiffusive CTRW mechanism and a regular random walk with Cauchy jumps as discussed here, we have used procedure described in \cite[Corollary 1]{sparre1953} and \cite{dybiec2009b} to check whether increments of the analyzed process are indeed statistically independent. Despite the fact that adapted displacement mechanism allows for trapping events due to rejection of some jumps (cf. Sec.~\ref{sec:model}), derived distributions of waiting times have not displayed slowly decaying long time asymptotics and duration of intervals between subsequent jumps stays in line with memoryless character of the process.

Properties of waiting time distribution and independence of increments fully confirm Markovianity of the asymptotic diffusive motion  of tracers meandering in a static, disordered monolayer of fibrinogens. We have therefore concluded that the observed sublinear MSD scaling is due to a  fractal structure of the environment in which the diffusion process takes place.

\begin{figure}[!htb]
\vspace{1cm}
\centerline{
\includegraphics[width=0.8\columnwidth]{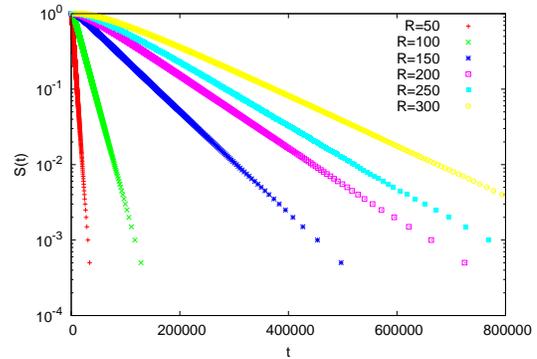}}
\vspace{1cm}
\centerline{
\includegraphics[width=0.8\columnwidth]{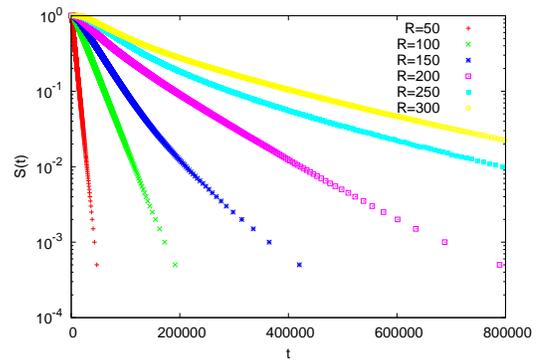}}
\caption{(Color online) Survival probability $S(t)$ for various radius $R=\{50,100,150,200,250,300\}$, i.e. the probability that at time $t$ a particle is still within a circle of given radius $R$ centered at its initial position, for ordered (top panel) and unordered (bottom panel) structures.
Jumps' length is drawn from the Cauchy distribution. The order parameter is $q=0.98$ (top panel) and $q=0.10$ (bottom panel).
}
\label{fig:arrivaltime}
\end{figure}

These findings have been further demonstrated in Figure~\ref{fig:arrivaltime} by using the concept of the survival probability $S(t)=\int_{|\vec{r}|\leqslant R}p(\vec{r},t)d^2\vec{r}$, i.e. the probability that at time $t$ a random walker starting its motion at a position $\vec{r}_0$ is still in the circle  of radius $R$ recorded for ordered (top panel) and unordered (bottom panel) monolayers. Various points correspond to various radii $R=\{50,\dots,300\}$. In the ordered case (top panel) the survival probability clearly displays exponential asymptotic for all values $R$, that is typical for Markovian processes. In contrast, in the unordered monolayer, the kinetics slows down and  escape times from the domain of radius $R$ become longer for large $R$. In consequence, the probability that a particle remains in the system up to a given time is significantly larger than in ordered monolayers and the tails of survival probability bend upwards, cf. bottom panel of Fig.~\ref{fig:arrivaltime}.

Moreover, due to coarse-graining of the space, the waiting time for the next jump are generated according to the geometric distribution (or exponential, for asymptotically continuous time problem) which is memoryless. Therefore, the studied process is a typical Markov chain with discrete number of states and transition rates, depending on the current position of the tracer.

Finally, we have checked if the observed processes can be differentiated using the multi-fractal detrended fluctuation analysis \cite{kantelhardt2002,makowiec2010,ihlen2012}. Nevertheless, such an analysis revealed only that observed trajectories are of the multi-fractal type. No systematic dependence of the multi-fractal quantifiers on the concentration, order parameter or jump length distribution have been observed (results not shown).

%
%
\section{Summary and Conlusions\label{sec:summary}}

The analysis of diffusion in crowded and limited space based on the mean squared displacement (MSD) demonstrates that the diffusion process is not very sensitive to the underlying random process generating jump lengths. In particular Gaussian random walk and L\'evy (Cauchy) flights results in processes of similar characteristics.
On the ensemble level, in both cases the subdiffusion characterized by the same effective exponents is observed. Therefore, presented result describing diffusion in crowded and anisotropic environment follow the pattern observed in \cite{ciesla2014}. 
The unified asymptotic behavior of the MSD is produced by obstacles which introduce effective cut-off to jump length distribution indicating a crucial role of truncation effects on properties of diffusive processes.
Nevertheless, short time properties clearly display difference between various jump length distributions.

Typically analysis of diffusion processes is based on examination of increments and waiting time distributions.
For the process in studied anisotropic medium the examination of increments and waiting times confirms Markovian character of the diffusion process in crowded environment.
Therefore, the sublinear scaling of the MSD is the consequence of the fractal structure of the environment. 
Finally, there is a big difference between ensemble derived and single trajectory derived properties of the diffusion process.
The MSD provides an ensemble based characteristics of the diffusion process which is not sufficient to discriminate between detailed properties of the displacement mechanism which are clearly visible on the single trajectory level. Consequently, analysis based on the MSD  can suggest that the process is subdiffusive despite its anomalously long increments.

%
%

\bibliography{bibliography}

\end{document}